\documentclass[sn-mathphys-num]{sn-jnl}

\usepackage{graphicx}%
\usepackage{multirow}%
\usepackage{amsmath,amssymb,amsfonts}%
\usepackage{amsthm}%
\usepackage{mathrsfs}%
\usepackage[title]{appendix}%
\usepackage{xcolor}%
\usepackage{textcomp}%
\usepackage{manyfoot}%
\usepackage{booktabs}%
\usepackage{algorithm}%
\usepackage{algorithmicx}%
\usepackage{algpseudocode}%
\usepackage{listings}%

\theoremstyle{thmstyleone}%
\theoremstyle{thmstyletwo}%

\theoremstyle{thmstylethree}%

\begin{document}

\title[Article Title]{Astrocytic NMDA Receptors Modulate the Dynamics of Continuous Attractors}


\author[1]{\fnm{Zihan} \sur{LIU}}

\author[2]{\fnm{Flavia Nathaline} \sur{CHANENTIA}}
\equalcont{These authors contributed equally to this work.}

\author[2]{\fnm{Patteera} \sur{SUPVITHAYANOND}}
\equalcont{These authors contributed equally to this work.}

\author*[3,4]{\fnm{Chi Chung Alan} \sur{FUNG}} \email{alan.fung@cityu.edu.hk}

\affil[1]{\orgdiv{Department of Computer Science}, 
\orgname{City University of Hong Kong}, 
\orgaddress{\street{Tat Chee Avenue}, \city{Kowloon Tong}, \state{Hong Kong}, \country{China}}}

\affil[2]{\orgdiv{Department of Biomedical Engineering}, 
\orgname{City University of Hong Kong}, 
\orgaddress{\street{Tat Chee Avenue}, \city{Kowloon Tong}, \state{Hong Kong}, \country{China}}}

\affil[3]{\orgdiv{Department of Neuroscience}, 
\orgname{City University of Hong Kong}, 
\orgaddress{\street{Tat Chee Avenue}, \city{Kowloon Tong}, \state{Hong Kong}, \country{China}}}

\affil[4]{\orgname{CityU Shenzhen Research Institute}, 
\orgaddress{\street{8 Yuexing 1st Road}, \city{Shenzhen Hi-tech Industrial Park, Nanshan District, Shenzhen}, \state{Guangdong}, \country{China}}}


\abstract{
Neuronal networking supports complex brain functions, with neurotransmitters facilitating communication through chemical synapses. The release probability of neurotransmitters varies and is influenced by pre-synaptic neuronal activity. Recent findings suggest that blocking astrocytic N-Methyl-D-Aspartate (NMDA) receptors reduces this variation. However, the theoretical implications of this reduction on neuronal dynamics have not been thoroughly investigated.
Utilizing continuous attractor neural network (CANN) models with short-term synaptic depression (STD), we explore the effects of reduced release probability variation. Our results show that blocking astrocytic NMDA receptors stabilizes attractor states and diminishes their mobility. These insights enhance our understanding of NMDA receptors' role in astrocytes and their broader impact on neural computation and memory, with potential implications for neurological conditions involving NMDA receptor antagonists.
 }




\maketitle

\section{Introduction}\label{sec1}

The intricate networking of neurons forms the foundation of complex brain functions and neural computations. At the heart of this communication lie neurotransmitters, which are released from pre-synaptic neurons and traverse chemical synapses to influence post-synaptic targets \cite{Luo2020-qp}. 

The probability of neurotransmitter release after pre-synaptic spikes is a critical factor in this process. The dynamics of neurotransmitter availability have been well described by the Tsodyks-Markram (TM) model \cite{Tsodyks1997}. Neurotransmitter release probability is not uniformly distributed across synapses \cite{Grillo2018,Jensen2021}. This variability in release probability plays a significant role in neural dynamics and synaptic plasticity.

Recent studies have highlighted the involvement of astrocytes, a type of glial cell, in modulating synaptic activity. Astrocytic N-Methyl-D-Aspartate (NMDA) receptors, in particular, have been shown to influence neurotransmitter release probabilities \cite{Chipman2021}. Blocking these receptors in astrocytes reduces the variation in release probabilities across synapses. Despite the novelty of these findings, the theoretical implications of reduced release probability variation on neuronal network dynamics remain largely unexplored. 

Moreover, NMDA receptor antagonists are used to control the symptoms of certain mental disorders \cite{Zorumski2015,Williams2016,Krystal2019}. However, since NMDA antagonists block not only neuronal NMDA receptors but also astrocytic NMDA receptors, understanding the consequences of blocking astrocytic NMDA receptors is particularly important \cite{Chipman2021}.

Continuous Attractor Neural Networks (CANNs) offer a powerful framework for investigating neural dynamics. CANNs are a family of neural network models that support continuous attractors \cite{Amari1977,Georgopoulos1993,BenYishai1995}. The network configuration is homogeneous along the attractor space \cite{Fung2010}. The attractor state is considered to represent continuous information in the nervous system, such as head direction \cite{Zhang1996}, self-location \cite{Battaglia1998}, and orientation of visual images \cite{BenYishai1995}. 

With short-term synaptic depression (STD), CANNs exhibit much richer dynamical behavior. STD results from the slow recovery of neurotransmitters after consumption, occurring over a time scale of hundreds of milliseconds \cite{Markram1998}. These rich dynamics include stability issues and the mobility of attractor states. Wang et al. (2015) demonstrated that CANNs with STD support chaotic behaviors \cite{Wang2015}. Understanding how variations in neurotransmitter release probability affect these dynamics is crucial for shedding light on the functional roles of astrocytic NMDA receptors.

In this study, we employ CANN models with STD to explore the effects of reduced release probability variation caused by blocking astrocytic NMDA receptors. We aim to elucidate how this reduction impacts the stability and mobility of attractor states within the network. Our findings provide significant insights into the role of NMDA receptors in astrocytes, with broader implications for neural computation, memory processes, and neurological conditions involving NMDA receptor antagonists.

\section{The Model} \label{sec2}

\subsection{Continuous Attractor Neural Networks (CANNs)}

We employ a continuous attractor neural network (CANN) model to investigate the effect due to the modulation of neurotransmitterr release probability. Early forms of CANNs appeared in 1977 \cite{Amari1977}. Other similar forms of CANNs addressing different brain functions were then proposed \cite{Georgopoulos1993,BenYishai1995,Stein2021}. In CANNs, neurons that have similar preferred stimuli are excitatory coupled, and there is a long-range inhibition as a balance (see Fig. \ref{fig:cann}(A)). Then, the network can support a {\it continuous} family of bump-shaped states for different preferred stimuli (see Fig. \ref{fig:cann}(B)). Those states are attractors of the network representing the information being encoded in the neural network. Those stimuli could be the animal's self-location \cite{Battaglia1998}, head direction\cite{Zhang1996}, or object orientation in the primary visual cortex \cite{BenYishai1995}.

In this study, there are $N$ excitatory neurons in the CANN. The dynamics of the neuronal input of neurons with preferred stimulus $x$ at time $t$, $u\left(x,t\right)$, evolves as \cite{Fung2012a}
\begin{equation}
\tau_s\frac{\partial u\left(x,t\right)}{\partial t}=-u\left(x,t\right)+\int dx^{\prime}J\left(x,x^{\prime}\right)p\left(x,x^{\prime},t\right)r\left[u\left(x,t\right)\right]+I^{\text{ext}}\left(x,t\right),\label{eq:dudt}
\end{equation}
where $\tau_s\sim 1~{\rm ms}$ is the time constant and $I^{\text{ext}}$ is the external input. $J\left(x,x^{\prime}\right)$ models the static coupling defined by
\begin{align}
J\left(x,x^{\prime}\right) & =\frac{J_{0}}{\sqrt{2\pi}a}\exp\left[-\frac{\left|x-x^{\prime}\right|^{2}}{2a^{2}}\right],\label{eq:Jxx}
\end{align}
where $J_{0} = 1$ is average strength of the static coupling and $a$ is the range of excitatory connections over the preferred stimulus space. $r$ is the neuronal activation after global inhibition given by
\begin{equation}
r\left(x,t\right)=\frac{u\left(x,t\right)^{2}}{1+\frac{k}{8\sqrt{2\pi}a}\int dx^{\prime}u\left(x^{\prime},t\right)^{2}},
\end{equation}
where $k$ controls the magnitude of divisive global inhibition \cite{Wu2005, Deneve1999}. 

\subsection{Short-term Synaptic Depression (STD)}

In Eq. (\ref{eq:dudt}), $p\left(x,t\right)$ models the modulation due to short-term synaptic depression of the pre-synaptic neurons having preferred stimulus $x$ at time $t$. It dynamics is defined by 
\begin{align}
\frac{\partial p\left(x,x^{\prime},t\right)}{\partial t} & =\frac{1-p\left(x,x^{\prime},t\right)}{\tau_{d}}-\beta\left(x,x^{\prime}\right)p\left(x,x^{\prime},t\right)r\left[u\left(x^{\prime},t\right)\right],\label{eq:dpdt}
\end{align}
where $\beta\left(x,x^{\prime}\right)$ is a magnitude for STD for the synapse connecting neurons at $x$ and $x^{\prime}$ and $\tau_{d} = 50 ms$ is the time constants corresponding to slow neurotransmitter recovery, which is between 25 ms and 100 ms \cite{Tsodyks1998}. In our previous studies, $\beta$ is assumed to be a constant across neurons and synapses \cite{Fung2012a}. 

\subsection{Generation of $\beta\left(x,x^{\prime}\right)$}

\begin{figure}
\begin{centering}
\includegraphics[width=13cm]{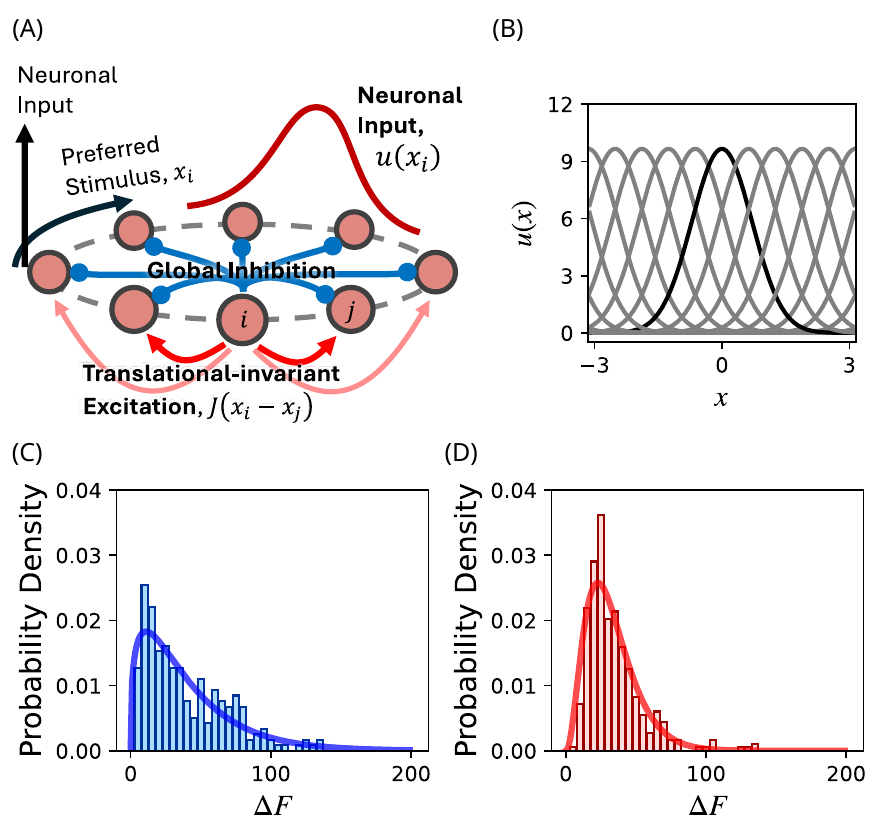}
\par\end{centering}
\caption{\label{fig:cann} Modeling with continuous-attractor neural networks (CANNs). (A) Illustration of local excitatory and global inhibitory connections in CANNs. The local excitatory connections are modeled by the translational-invariant excitatory connection $J(x-x^\prime)$. This configuration supports localized activity profiles, i.e., bumps, on the preferred stimulus space. (B) A family of bump-shaped states of a CANN encoding different stimuli over the preferred stimulus space. Parameters: $N=100$, $k=0.5$, $a=0.5$ and $\beta(x,x^\prime)=0$. (C) Distribution of FM1-43 signals ($\Delta F$) under the control condition. (D) Distribution of $\Delta F$ where astrocytic NMDA receptors were blocked. Both (C) and (D) can be fitted by Gamma distributions. These gamma distributions were used to determine the release probabilities since $\Delta F$ represents the readily releasable pool of neurotransmitters. Fitted parameters for (C): $\kappa = 1.378$ and $\theta = 29.196$. Fitted parameters for (D): $\kappa = 3.355$ and $\theta = 9.744$.}
\end{figure}

Because $\beta\left(x,x^{\prime}\right)$ is the characteristic consumption rate of neurotransmitters that is proportional to the release probability, to generate $\beta\left(x,x^{\prime}\right)$ for our study, we first generate a number of gamma random numbers,
\begin{equation}
\beta\left(x,x^{\prime}\right)\sim f\left(x;\kappa,\theta\right)=\frac{1}{\Gamma\left(\kappa\right)\theta^{\kappa}}x^{\kappa-1}e^{-\frac{x}{\theta}},\label{eq:beta_gamma}
\end{equation}
where $\kappa$ and $\theta$ are the shape parameter and the scaling parameter respectively. The parameters are derived by fitting gamma distributions to the data \cite{Chipman2021}, as depicted in panels (C) and (D) of Figure \ref{fig:cann}. These figures illustrate the distributions of FM1-43 signals ($\Delta F$) observed in two scenarios: the control condition (panel (C)), and the scenario where astrocytic NMDA receptors were blocked (panel (D)). Since the FM1-43 signal represents the readily releasable pool of neurotransmitters \cite{Cochilla1999}, we assume that the gamma distributions also apply to release probabilities. In the following investigation, parameter $\overline{\beta}$ controls the mean magnitude of $\beta\left(x,x^{\prime}\right)$. Conditions represented by Figs. \ref{fig:cann} (C) and (D) will be investigated in this study. Additionally, considering that release probability and synaptic connection strength are inversely proportional to the distance between the cell body and the connecting site \cite{Grillo2018,Jensen2021}, we posit that the release probability is directly related to the synaptic connection strength. Subsequently, the pre-generated random numbers are sorted, and synapses are assigned random numbers based on their synaptic connection strength.

\section{Results}\label{sec3}

\subsection{Profiles of $\text{\ensuremath{\beta}\ensuremath{\left(x,x^{\prime}\right)}}$
and $p\left(x,x^{\prime},t\right)$}

\begin{figure}
\begin{centering}
\includegraphics[width=13cm]{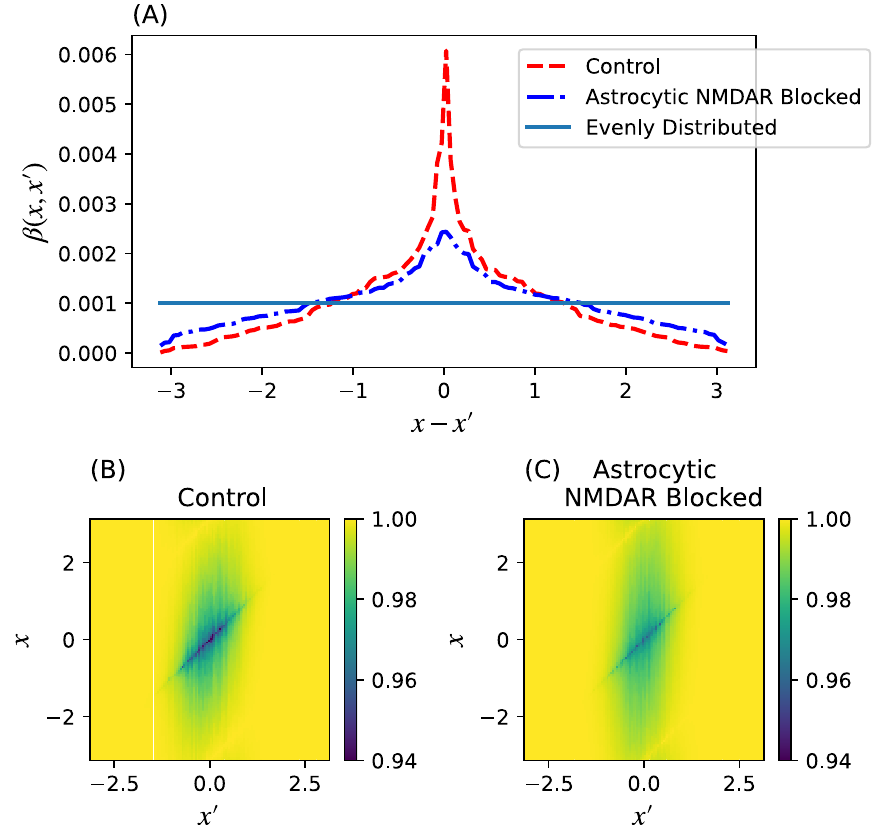}
\par\end{centering}
\caption{\label{fig:beta_n_pxx}
Profiles of $\beta\left(x,x^{\prime}\right)$ and $p\left(x,x^{\prime},t\right)$ under different conditions. (A) Profile of $\beta$ as a function of $\left(x-x^{\prime}\right)$, note that $\beta\left(x,x^{\prime}\right)$ is proportional to the release probability of neurotransmitters.   
(B) Profile of $p\left(x,x^{\prime},t\right)$ under the control condition. (C) Profile of $p\left(x,x^{\prime},t\right)$ under astrocytic NMDAR blocked condition. Parameters: $N=128$, $a=0.5$ and $\overline{\beta}=0.001$. }
\end{figure}

In the setting of the current model, the magnitude of $\beta$ of
the synapse connecting neurons at $x$ and $x^{\prime}$depends on
$\left(x-x^{\prime}\right)$. According to the work by Grillo et al
(2018), and Jensen et al (2021), the release probability of neurotransmitters
on a synapse should be proportional to synaptic weight \cite{Grillo2018,Jensen2021}.
Although the neurotransmitter is log-normally distributed (see Figs.
\ref{fig:cann} (C) and (D)), after reordering the release probabilities
by synaptic weights, the neurotransmitter release probability will
also be a function of $\left(x-x^{\prime}\right)$, as shown in Fig.
\ref{fig:beta_n_pxx}(A). In the control condition, the maximum release
probability is higher than that of the astrocytic NMDAR blocked condition,
even though the mean release probability is shared by both conditions. 

The reduction of variation of release probabilities caused by blocking
astrocytic NMDAR can modulate not only $\beta\left(x,x^{\prime}\right)$
but also $p\left(x,x^{\prime},t\right)$. Suppose the system described
by Eqs. (\ref{eq:dudt}) and (\ref{eq:dpdt}) reached its equilibrium
state and the state is static. Then, we have
\begin{align}
p\left(x,x^{\prime},t\right) & =\frac{1}{1+\tau_{d}\beta\left(x,x^{\prime}\right)r\left[u\left(x^{\prime},t\right)\right]}\\
 & \approx1-\tau_{d}\beta\left(x,x^{\prime}\right)r\left[u\left(x^{\prime},t\right)\right]\text{ , if }\max\beta\left(x,x^{\prime}\right)\text{ is small.}\label{eq:pxx_static_small}
\end{align}
The profiles of $p\left(x,x^{\prime},t\right)$ from different conditions
obtained from simulations are shown in Fig. \ref{fig:beta_n_pxx}
(B) and (C). These results agree with the prediction by Eq. (\ref{eq:pxx_static_small}).
By comparing the $p$ profiles shown in Fig. \ref{fig:beta_n_pxx}
(B) and (C), there is a difference between two conditions, even though
they shared the same $\overline{\beta}$. In the control condition,
the minimum of $p\left(x,x^{\prime}\right)$ is smaller. However,
in the astrocytic NMDAR blocked condition, the $p$ profile has a
wider range. This difference is consistent with the results for $\beta\left(x,x^{\prime}\right)$shown
in Fig. \ref{fig:beta_n_pxx}(A).

\subsection{Translational Stability Analysis}

One important effect of short-term synaptic depression (STD) on continuous-attractor
neural networks (CANNs) is the transitional destabilization of the
network states in the attractor space. To investigate the translational
stability, we may check existence of the intrinsic motion solution
for a given parameter set in the absence of external input, i.e., $I^{\rm ext} = 0$. To beginning with, we consider 
\begin{align}
u\left(x,t\right) & =u_{0}\left(t\right)E_{4}\left(x-z\left(t\right)\right)\label{eq:ux_static}\\
p\left(x,x^{\prime},t\right) & =1-p_{0}\left(t\right)\psi_{0}\left(x-z(t)-s\left(t\right),x^{\prime}-z(t)-s\left(t\right)\right)\label{eq:px_static}
\end{align}
where $z\left(t\right)$ is the center of $u\left(x,t\right)$, $E_{4}=\exp\left[-\left|x\right|^{2}/ (4a^{2})\right]$.
In particular, 

\begin{equation}
\psi_{0}\left(x,x^{\prime}\right)\sim\tilde{\beta}\left(x,x^{\prime}\right)E_{2}\left(x^{\prime}\right),\label{eq:psi}
\end{equation}
where and $E_{2}=\exp\left[-\left|x\right|^{2}/ (2a^{2})\right]$
and $\tilde{\beta}$ is a gamma random number sorted according
to the synaptic weight with a unity mean. $s\left(t\right)$ is the
separation between the $u$ profile and the $p$ profile. One should
note that the shape and scaling parameters should be the same as the
gamma distribution generating $\beta\left(x,x^{\prime}\right)$ in
Eq. \ref{eq:beta_gamma}. By substituting Eqs. (\ref{eq:ux_static})
and (\ref{eq:px_static}) into Eqs. (\ref{eq:dudt}) and (\ref{eq:dpdt}),
we have
\begin{align}
\frac{du_{0}\left(t\right)}{dt}E_{4}\left(x-z\left(t\right)\right)-\frac{dz}{dt}u_{0}\left(t\right)\frac{dE_{4}}{dx}\left(x-z\left(t\right)\right) & =F_{u}\left(u,p\right) \label{eq:dudt_perturb_1}\\
-\frac{dp_{0}\left(t\right)}{dt}\psi_{0}+p_{0}\left(t\right)\left[\frac{dz}{dt}+\frac{ds}{dt}\right]\left(\frac{\partial\psi}{\partial x}+\frac{\partial\psi}{\partial x^{\prime}}\right) & =F_{p}\left(u,p\right) \label{eq:dpdt_perturb_1}
\end{align}
Here $\psi_{0}$ and its derivatives are evaluated at $\left(x-z(t)-s\left(t\right),x^{\prime}-z(t)-s\left(t\right)\right)$.
At the steady state, we set $\frac{du_{0}\left(t\right)}{dt}$, $\frac{dp_{0}\left(t\right)}{dt}$
and $\frac{ds}{dt}$ equal to zero. By projecting even and odd functions,
we can obtain the following equations.

\begin{align}
\int dxE_{4}\left(x-z\left(t\right)\right)F_{u}\left(u,p\right) & =0\label{eq:pert_E4Fu}\\
-\frac{\int dx\frac{dE_{4}}{dx}\left(x-z\left(t\right)\right)F_{u}\left(u,p\right)}{\int dx\left[\frac{dE_{4}}{dx}\left(x-z\left(t\right)\right)\right]^{2}} & =u_{0}\frac{dz}{dt}\label{eq:pert_dE4dxFu}\\
\int dx\int dx^{\prime}\psi_{0}F_{p}\left(u,p\right) & =0\label{eq:pert_phiFp}\\
\frac{\int dx\int dx^{\prime}\left(\frac{\partial\psi}{\partial x}+\frac{\partial\psi}{\partial x^{\prime}}\right)F_{p}\left(u,p\right)}{\int dx\int dx^{\prime}\left(\frac{\partial\psi}{\partial x}+\frac{\partial\psi}{\partial x^{\prime}}\right)^{2}} & =p_{0}\frac{dz}{dt}\label{eq:pert_dphidxFp}
\end{align}
The fixed point solutions for $u_{0}$, $p_{0}$ and $s$, and $\frac{dz}{dt}$
can be solved by numerical methods. The fixed point solution can be
used to deduce how the translational destabilization could take place
for different different $k$ and $\overline{\beta}$. By solving,
if $\frac{dz}{dt}=0$ is the only solution, then the system with that
combinations of $k$ and $\overline{\beta}$ supports only static
solutions. However, if $\frac{dz}{dt}\ne0$ is the one of the solutions,
then the CANN is able to support traveling $u$ profiles and $p$
profiles. 

\begin{figure}
\begin{centering}
\includegraphics[width=13cm]{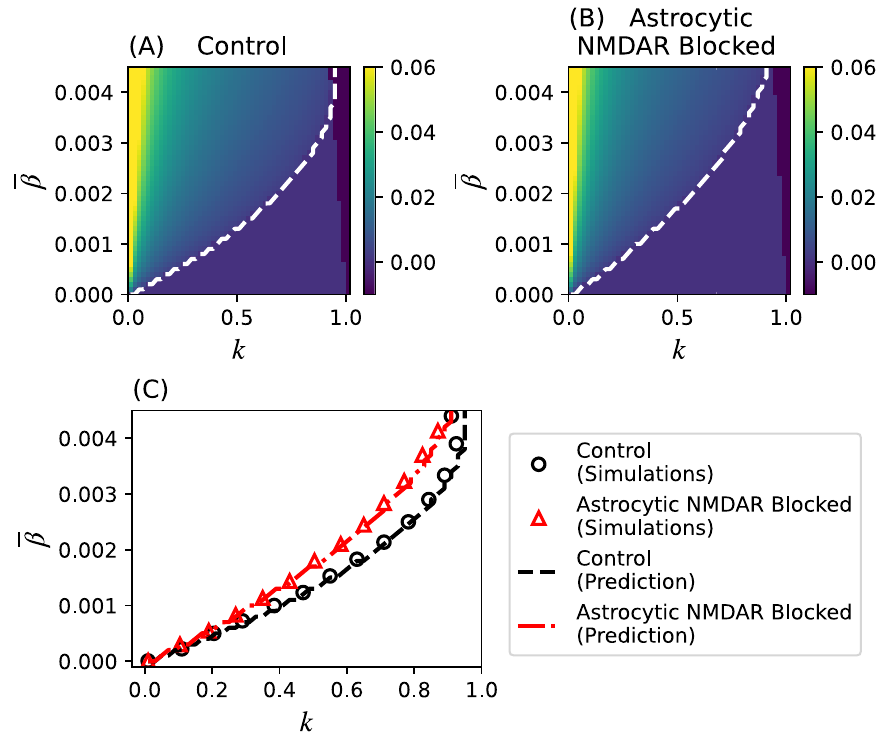}
\par\end{centering}
\caption{\label{fig:std_boundary}
 Phase-plane diagrams for analyzingg state transition. 
 (A) The color code shows the speed of the spontaneous motion measurements of CANNs with STD under the control condition, where the white curve represents the boundary obtained by perturbative analysis. The left side of the curve represents the spontaneous motion phase (speed larger than zero), and the right side represents the static phase (speed equals zero). On the right edge of the color map, there is a narrow region with negative speed. That negative-speed region represents the regions supporting the trivial solution only, i.e., $u = 0$.
 (B) Same as (A) but under the astrocytic NMDAR blocked condition. 
 (C) Comparisons of spontaneous speed between simulations and prediction under different conditions. The results show that under control conditions, smaller $\overline{\beta}$ is able to destabilize the attractor state.
 Parameters: $N=128$ and $a=0.5$. 
 }
\end{figure}

In Figs. \ref{fig:std_boundary}(A) and (B), there are boundaries
given by the above perturbative analysis (the white curves) plotted
along with the simulations (the color map) for different conditions
(control condition and astrocytic NMDAR blocked condition). The area
in the color map with positive values means the non-zero speed of spontaneous
motion ($\frac{dz}{dt}\ne0$). The area in the color map with zero
value means static profiles. The area in the color map with negative
values means the trivial solution is the only solution, i.e. $u\left(x,t\right)=0$
and $p\left(x,x^{\prime},t\right)=1$. The nice agreement between
the prediction and simulations in the phase transition boundary suggested
that the misalignment between $u$ profile and $p$ profiles plays
a vital role in translational destabilization. 

In Fig. \ref{fig:std_boundary}(C), despite the nice agreement between
simulations and predictions, there is a notable difference in boundaries
between the control condition and the astrocytic NMDAR blocked scenario.
The result suggests that in the control condition, a smaller $\overline{\beta}$
will be needed to destabilize the attractor states. It implies that
blocking astrocytic NMDAR can result in stabilizing the attractor
if the parameter of the CANNs with STD is near the boundary separating
static phase and spontaneous motions. 

\subsection{Release Probability Variation Reduction Modulates Spontaneous Motion}

The equation system shown in Eqs. (\ref{eq:pert_E4Fu}) - (\ref{eq:pert_dphidxFp})
can show not only the modulation of the transition boundaries between
static phase and spontaneous moving phase (Fig. \ref{fig:std_boundary}(C)),
but also the speed of spontaneous motions supported by short-term
synaptic depression. 

\begin{figure}
\begin{centering}
\includegraphics[width=13cm]{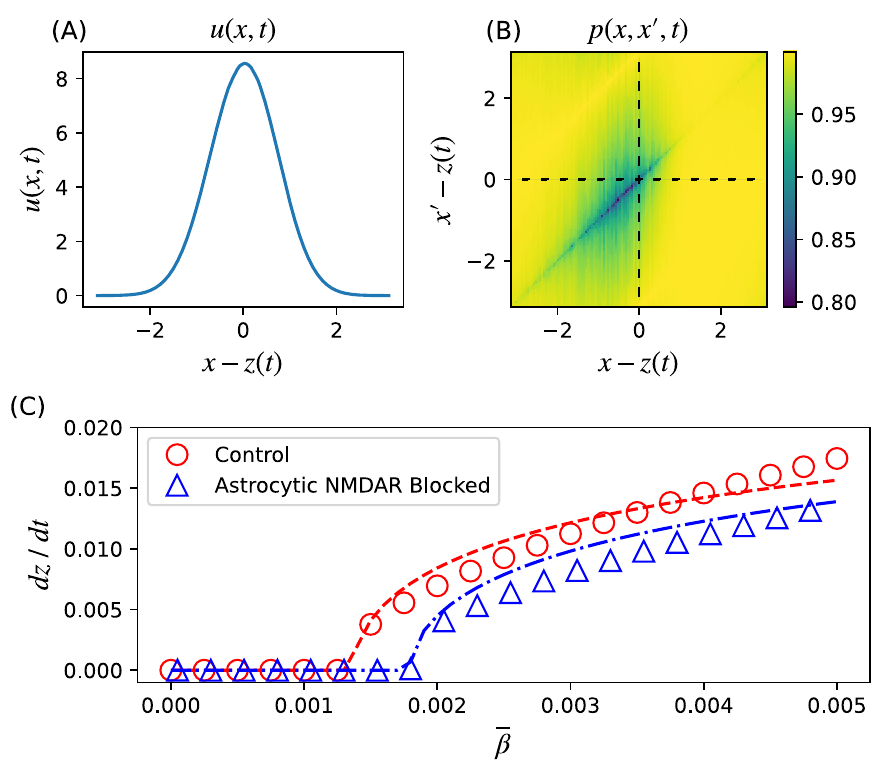}
\par\end{centering}
\caption{\label{fig:moving_and_speed}
Analysis for spontaneous motion. (A) Snapshot of $u$ profile in its moving phase. (B) Snapshot of $p$ profile in its moving phase. Dashed lines represent the center of mass of $u$ in the $x$ and $x^\prime$ coordinates. Comparison between (A) and (B) shows the misalignment between $u$ profile and $p$ profile in a simulation. Parameters: $N=128$, $k=0.5$, $\overline{\beta} = 0.00003$ and $a=0.5$. (C) Blocking Astrocytic NMDAR degrades the spontaneous motion speed. Parameters: $N=128$, $k=0.5$ and $a=0.5$.
}
\end{figure}

In the perturbative analysis, we assume the misalignment between $u$
profile and $p$ profile in the preferred stimulus space is the direct
cause supporting the spontaneous motion. To show this in simulations,
we have snapshotted the $u$ profile and $p$ profile in a simulation.
In Fig. \ref{fig:moving_and_speed}(A), there is a snapshot of $u$
profile in its moving phase. The $u$ profile shows that the shape
is not significantly modulated. However, the magnitude of the $u$
profile became shorter than CANN without the influence of short-term
synaptic depression (\textit{c.f.} Fig. \ref{fig:cann}(B)). For the
$p$ profile, the position relative to the center of the $u$ profile
($z\left(t\right)$) is misaligned (Fig. \ref{fig:moving_and_speed}(B)).
The result is consistent with the previous report \cite{Fung2012a}. 

By investigating the spontaneous speed in detail, we can see that
the modulation in release probability variation caused by blocking
astrocytic NMDA receptors affect not only the transition boundary.
It also degrades the spontaneous moving speed even though the same
$\overline{\beta}$ could destabilize both conditions. This result
coincides with the results in Fig. \ref{fig:std_boundary}(C) that
blocking astrocytic NMDA receptors could result in an enhanced translational
stability.

\subsection{Blocking Astrocytic NMDA Receptors Slow Down the Tracking to a Stimulus}

In addition to the intrinsic property, the responses of CANNs to external input defines the functional meaning of the model. In our previous studies, the network state will change its position to respond to the change of external input. Here, the external input is defined by 
\begin{align}
I^{\rm ext}\left(x,t\right) &= A \times \exp \left[-\frac{\left|x-z_0(t)\right|^2}{4a^2}\right]    ,
\end{align}
where $A$ is the magnitude of the external input and $z_0$ is the position the external stimulus. The passage time of the tracking dynamics can be considered as the reaction time of the network to catch up the abrupt shift of the external input\cite{Wu2008, Fung2010}. In the version of CANNs without STD, the reaction time is similar to a logarithmic function. Short-term synaptic depression (STD) can shorten the reaction time by destabilizing the neuronal activity state \cite{Fung2012a}. However, the previous study investigated the scenario with uniformly distributed release probability in CANNs. Here, we investigated the reaction time of CANNs with STD under the control and astrocytic NMDAR-blocked conditions. 

\begin{figure}
\begin{centering}
\includegraphics[width=13cm]{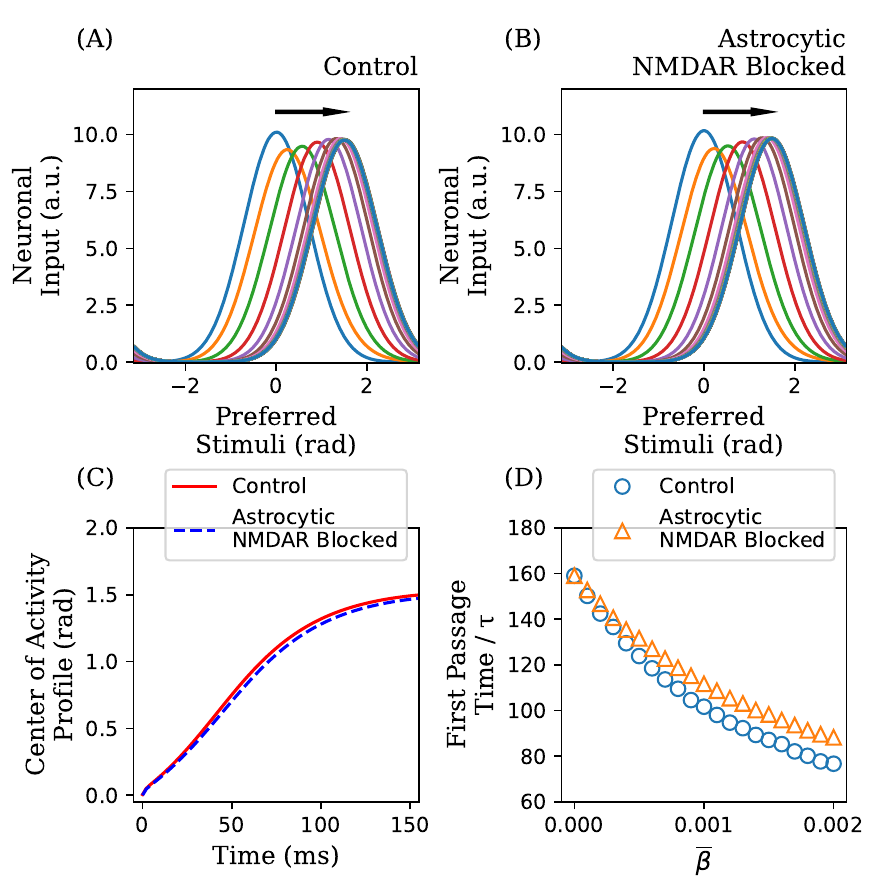}
\par\end{centering}
\caption{\label{fig:tracking}
Blocking Astrocytic NMDA Receptors Modulates the Tracking Dynamics.
(A) Snapshots of the network input states to respond to the change of the stimulus position from 0 to 1.5 in the control condition.  Parameters: $N=128$, $k=0.5$, $\overline{\beta} = 0.0005$, $a=0.5$ and $A = 0.5$.
(B) Same as (A) but in the condition that astrocytic NMDA receptors are blocked.
(C) Trajectories of the centers of mass of the neuronal input shown in (A) and (B).
(D) The first passage times of the neuronal input profiles shown in (A) and (B). 
}
 
\end{figure}

Figures \ref{fig:tracking} (A) and (B) show snapshots of the tracking progress of $u(x,t)$. These results look similar. However, the tracking process under the control condition is faster, while the tracking process became slower under the astrocytic NMDAR-blocked condition, see Fig. \ref{fig:tracking}(C). The corresponding passage time is shown in Fig. \ref{fig:tracking} (D). This result qualitatively agrees with the above results. Figures \ref{fig:moving_and_speed} and \ref{fig:std_boundary} suggested blocking astrocytic NMDAR will reduce the translational destabilization of CANN attractor states. A slower tracking processing is consistent with a less-destabilized CANN state. 

\subsection{Noise Response of Attractor States}

In addition to the response to the change of external stimulus shown in Fig. \ref{fig:tracking}, the response of the network state to external noise is also of interest. To perform such an investigation, we modified Eq. (\ref{eq:dudt}) to incorporate the noise term. 
\begin{equation}
\tau\frac{\partial u\left(x,t\right)}{\partial t}=-u\left(x,t\right)+\int dx^{\prime}J\left(x,x^{\prime}\right)p\left(x,x^{\prime},t\right)r\left[u\left(x,t\right)\right]+\eta\left(x,t\right),\label{eq:dudt_noise}
\end{equation}
Here $\eta\left(x,t\right)$ is a white noise given that $\langle \eta\left(x,t\right)\eta\left(x^\prime,t^\prime\right)\rangle = 2T \delta\left(t-t^\prime\right)\delta\left(x-x^\prime\right)$. $T$ is the noise temperature. 

In a CANN without STD, the perturbative analysis suggests the drifting speed of the network state should be proportional to white noise in the temporal domain. The drifting of the network state should be a Brownian motion along the preferred stimulus space. Here, we expect the drifting of the network states in a CANN in the current study should also be a Brainian motion, i.e., $\langle\Delta z\left(t\right)^2\rangle$ should be proportional to time.

To investigate the noise response, we adopt the perturbative analysis similar to Eqs. (\ref{eq:dudt_perturb_1}) and (\ref{eq:dpdt_perturb_1}).
\begin{align}
\frac{du_{0}\left(t\right)}{dt}E_{4}\left(x-z\left(t\right)\right)-\frac{dz}{dt}u_{0}\left(t\right)\frac{dE_{4}}{dx}\left(x-z\left(t\right)\right) & =F_{u}\left(u,p\right) +\eta\left(x,t\right) \label{eq:dudt_perturb_2}\\
-\frac{dp_{0}\left(t\right)}{dt}\psi_{0}+p_{0}\left(t\right)\left[\frac{dz}{dt}+\frac{ds}{dt}\right]\left(\frac{\partial\psi}{\partial x}+\frac{\partial\psi}{\partial x^{\prime}}\right) & =F_{p}\left(u,p\right) .\label{eq:dpdt_perturb_2}
\end{align}
Since the drifting speed is coupled with the separation between the centers of mass of $u\left(x,t\right)$ and $1-p\left(x,x^\prime,t\right)$, we consider only $z(t)$ and $s(t)$. Then we have
\begin{align}
    \frac{dz}{dt} &= M_{zs}s-\tilde{\eta}\left(t\right) \\
    \frac{ds}{dt} &= M_{ss}s+\tilde{\eta}\left(t\right),
\end{align}
where
\begin{align}
    \tilde{\eta}\left(t\right) &=\frac{1}{u_{0}\int dx\left(\frac{dE_{4}}{dx}\right)^{2}}\int dx\frac{dE_{4}}{dx}\eta\left(x,t\right) \\
    M_{zs} &= -\frac{1}{u_{0}\int dx\left(\frac{dE_{4}}{dx}\right)^{2}}\int dx\frac{dE_{4}}{dx}\frac{\partial F_{u}}{\partial s} \\
    M_{ss} &= \frac{1}{p_{0}\int dx\int dx^{\prime}\left(\frac{\partial\psi}{\partial x}+\frac{\partial\psi}{\partial x^{\prime}}\right)^{2}}\int dx\int dx^{\prime}\left(\frac{\partial\psi}{\partial x}+\frac{\partial\psi}{\partial x^{\prime}}\right)\frac{\partial F_{p}}{\partial s} \nonumber \\
    & ~~~~~~~~+\frac{1}{u_{0}\int dx\left(\frac{dE_{4}}{dx}\right)^{2}}\int dx\frac{dE_{4}}{dx}\frac{\partial F_{u}}{\partial s}
\end{align}
By calculating, we have 
\begin{equation}
    \langle \Delta z \left( t \right)^2\rangle = 
    \frac{4\sqrt{2}a}{u_{0}^{2}\sqrt{\pi}}T\left(1+\frac{M_{zs}}{M_{ss}}\right)^{2} \times t \equiv Dt.
    \label{eq:drift}
\end{equation}
Here, $D$ is the diffusion constant in Brownian motion. Detailed derivation can be found in the supplementary material.

\begin{figure}
\begin{centering}
\includegraphics[width=13cm]{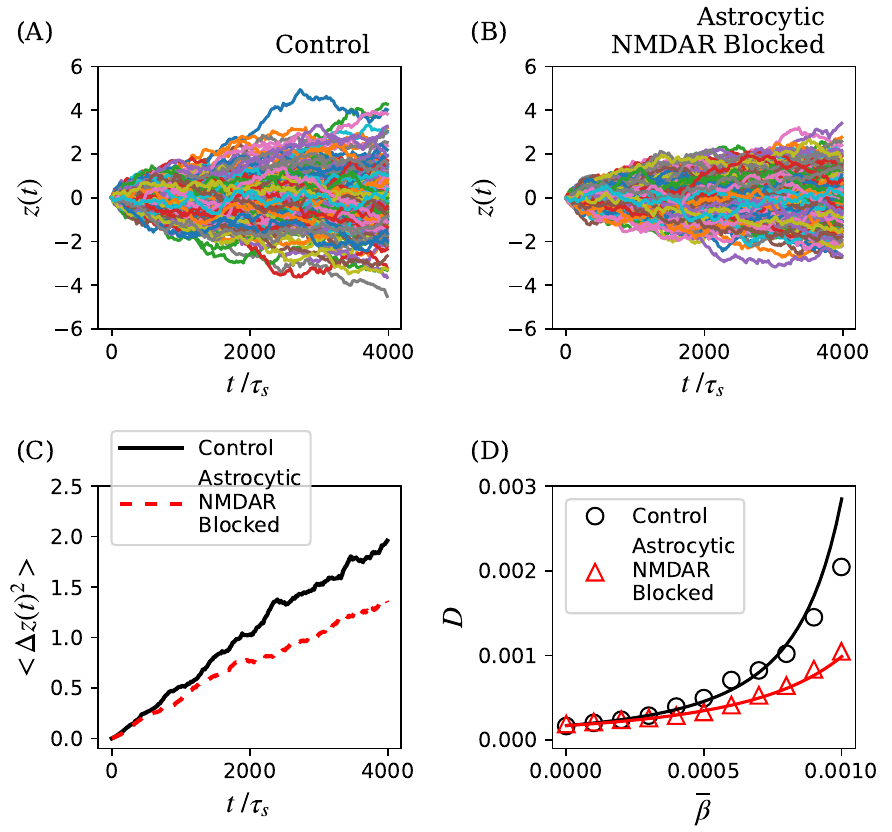}
\par\end{centering}
\caption{\label{fig:noise}
Blocking astrocytic NMDA receptors Reduces the Noise Responses of Attractor States. 
(A) 200 samples of center of mass of neuronal inputs under the influence of white noise in the control condition. Parameters: $N=128$, $k=0.5$, $\overline{\beta} = 0.0005$, $a=0.5$ and $T = 0.01$.
(B) same as (A) but in the condition that astrocytic NMDA receptors are blocked.
(C) Mean-square of center of mass of neuronal inputs under different conditions.
(D) The measured diffusion constants versus $\overline{\beta}$ in different conditions. Symbols: simulations. Curves: predictions by Eq. (\ref{eq:drift}).
}

\end{figure}

The prediction given by Eq. (\ref{eq:drift}) suggested that the drifting of the network state will be similar to CANNs without STD under the influence of white noise. As shown in Figs. \ref{fig:noise}(A) and (B), the centers of mass of the network states diffused from its original positions under the influence of white noise. One may notice that the diffusion for the astrocytic NMDAR blocked case, the variation of the diffusion across realization, is smaller. It is consistent with the above results that blocking astrocytic NMDAR can reduce the mobility of network states. A more comprehensive comparison can be found in Fig. \ref{fig:noise}(C). The mean-square of the drifting displacement among simulations is approximately a linear function of time, which agrees with the prediction by Eq. (\ref{eq:drift}). On the other hand, the average STD magnitude $\overline{\beta}$ can also modulate the slope shown in Fig. \ref{fig:noise}(C). Figure \ref{fig:noise}(D) suggests a larger $\overline{\beta}$ can result in a larger diffusion constant. Also, blocking astrocytic NMDAR can reduce diffusion of attractor states. All these results are consistent with the above results and confirm that blocking astrocytic NMDAR can stabilize the network attractor states.

\section{Discussion}\label{sec4}

Our investigation was driven by recent experimental findings indicating that blocking N-Methyl-D-Aspartate (NMDA) receptors on astrocytes significantly reduces the variability of neurotransmitter release probabilities across neurons \cite{Chipman2021}.

Using a continuous attractor neural network (CANN) model, we implemented the phenomenon of release probability distribution narrowing and examined its effects on neural network dynamics. Our results show that blocking astrocytic NMDA receptors narrows the distribution of neurotransmitter release probabilities, leading to a more uniform release probability across different synapses of a neuron. This modulation has profound effects on the stability and mobility of attractor states within the CANN. Specifically, we found that the reduced variation in release probabilities stabilizes the attractor states, characterized by slower tracking dynamics (see Fig. \ref{fig:tracking}) and smaller noise response (see Fig. \ref{fig:noise}).

The enhanced stability of attractor states results in reduced mobility within the network, meaning that the attractor states are less likely to shift or move in response to external fluctuations. This suggests a more rigid representation of continuous information, potentially impacting how information is processed and maintained in the brain. Our findings indicate that the modulation of release probability by astrocytic NMDA receptors plays a crucial role in neural computation, regulating the stability and mobility of attractor states, which may influence cognitive functions reliant on stable neural representations.

Our study extends the theoretical understanding of CANNs by incorporating the variability in release probabilities and its modulation. These findings highlight the need for further experimental investigations to validate the theoretical predictions and explore the broader implications of astrocytic NMDA receptor blockage on neural dynamics.

Blocking NMDA receptors is a strategy used to treat mental disorders. NMDA antagonists are a class of drugs that inhibit the activity of NMDA receptors, a subtype of glutamate receptors in the brain. These receptors play a critical role in synaptic plasticity, cognitive functions, and neurotransmission \cite{Paoletti2013}. NMDA antagonists have garnered significant attention in the medical field due to their potential therapeutic applications. They are used in the treatment of various neurological and psychiatric disorders, including Alzheimer's disease, major depressive disorder, and certain types of chronic pain \cite{Parsons1998, Zarate2006}. By blocking NMDA receptors, these drugs can modulate excessive excitatory neurotransmission, which is often associated with neurodegenerative diseases and conditions characterized by neuronal overactivity \cite{Hardingham2010}. Despite their therapeutic potential, the use of NMDA antagonists must be carefully managed due to the risk of side effects, including cognitive impairment and neurotoxicity, making ongoing research into their mechanisms of action and optimal clinical use essential \cite{Muir1995}.

The results of the present study suggest a nontrivial implication that NMDA antagonist intake may influence the mobility of attractor states. The reduction in mobility of attractor states may degrade the ability of neural anticipation supported by short-term synaptic depression \cite{Fung2012b}. This implication suggests a potential side effect of NMDA antagonists when used as medication, warranting further investigation in the future. Moreover, working memory can be investigated using attractor models \cite{Stein2021}. The present study also implies a potential benefit of NMDA antagonists in treating mental disorders such as Major Depressive Disorder (MDD) \cite{Zarate2006} and Bipolar Disorder \cite{Diazgranados2010}.

However, there are limitations to the current study. Firstly, CANNs are abstract models best suited for investigating working memories. More comprehensive neuronal network models will be needed to study other implications of blocking astrocytic NMDA receptors. Secondly, the detailed mechanism of the release probability narrowing process is still unclear. There may be other factors to consider when studying the dynamics. Lastly, blocking astrocytic NMDA receptors without affecting neuronal NMDA receptors in living animals is still challenging. New experimental techniques will be needed to validate the present theoretical study.

\section{Conclusion}\label{sec5}


In this study, we investigated the role of astrocytic NMDA receptors in modulating neurotransmitter release probability and the dynamics of continuous attractor neural networks (CANNs) with short-term synaptic depression (STD). Our findings indicate that blocking astrocytic NMDA receptors stabilizes attractor states within neural networks, leading to diminished mobility of these states. This stabilization effect enhances our understanding of the complex interplay between astrocytes and neurons, particularly how astrocytic NMDA receptor activity influences synaptic transmission and neural computation.

The implications of these findings are significant for both basic neuroscience and clinical applications. By elucidating the mechanisms through which astrocytic NMDA receptors regulate synaptic variability, we provide a foundation for further exploration into how these processes affect memory and learning. 

Future research could extend these insights by exploring the specific pathways through which astrocytes impact other types of synaptic plasticity and network dynamics. Overall, our study contributes to a deeper understanding of astrocyte-neuron interactions and their broader implications for brain function and disease.

\backmatter

\section*{Supplementary information}

\subsection*{Noise Response Analysis}

For simpicity, let us define 
\begin{align}
F_{u}\left[u,p\right] & =\frac{1}{\tau_{s}}\left\{ -u\left(x,t\right)+\int dx^{\prime}J\left(x,x^{\prime}\right)p\left(x,x^{\prime},t\right)r\left[u\left(x,t\right)\right]+\eta\left(x,t\right)\right\} \\
F_{p}\left[u,p\right] & =\frac{1}{\tau_{d}}\left\{ 1-p\left(x,x^{\prime},t\right)-\beta\left(x,x^{\prime}\right)p\left(x,x^{\prime},t\right)r\left[u\left(x^{\prime},t\right)\right]\right\} 
\end{align}
To begine with, let us assume the functional form of $u$ and $p$
is almost preserved like its static states: 

\begin{align}
u\left(x,t\right) & =u_{0}\left(t\right)E_{4}\left(x-z\left(t\right)\right)\\
p\left(x,x^{\prime},t\right) & =1-p_{0}\left(t\right)\psi\left(x-z\left(t\right)-s\left(t\right),x^{\prime}-z\left(t\right)-s\left(t\right)\right)
\end{align}
where $z\left(t\right)$ is the center of mass of $u$, $E_{4}\left(x\right)=\exp\left[-x^{2}/\left(4a^{2}\right)\right]$,
$s\left(t\right)$ is the misalignment (or separation) between the
$u$ profile and the $p$ profile, and

\begin{equation}
\psi_{0}\left(x,x^{\prime}\right)\sim\beta\left(x,x^{\prime}\right)E_{2}\left(x^{\prime}\right).
\end{equation}
$u_{0}\left(t\right)$ and $p_{0}\left(t\right)$ are dynamical variables
for the variations of the magnitudes of $u$ and $1-p$. In particular,
\begin{align}
\frac{\partial\psi}{\partial x}+\frac{\partial\psi}{\partial x^{\prime}} & =\frac{\partial\beta\left(x,x^{\prime}\right)}{\partial x}E_{2}\left(x^{\prime}\right)+\frac{\partial\beta\left(x,x^{\prime}\right)}{\partial x^{\prime}}E_{2}\left(x^{\prime}\right)+\beta\left(x,x^{\prime}\right)\frac{\partial E_{2}\left(x^{\prime}\right)}{\partial x^{\prime}}\nonumber \\
 & =\beta\left(x,x^{\prime}\right)\frac{\partial E_{2}\left(x^{\prime}\right)}{\partial x^{\prime}}\text{, since }\beta\left(x,x^{\prime}\right)=\beta\left(x-x^{\prime}\right)
\end{align}
Here, one should also note that $E_{2}=\exp\left[-x^{2}/\left(2a^{2}\right)\right]$.
For $u\left(x,t\right)$, we have
\begin{equation}
\frac{\partial u\left(x,t\right)}{\partial t}=\frac{du_{0}\left(t\right)}{dt}E_{4}\left(x-z\left(t\right)\right)-\frac{dz}{dt}u_{0}\left(t\right)\frac{dE_{4}}{dx}\left(x-z\left(t\right)\right).
\end{equation}
For $p\left(x,x^{\prime},t\right)$, we have
\begin{align}
\frac{\partial p\left(x,x^{\prime},t\right)}{\partial t}= & -\frac{dp_{0}\left(t\right)}{dt}\psi\left(x-z\left(t\right)-s\left(t\right),x^{\prime}-z\left(t\right)-s\left(t\right)\right)\nonumber \\
 & +p_{0}\left(t\right)\left(\frac{dz}{dt}+\frac{ds}{dt}\right)\frac{\partial\psi}{\partial x}\left(x-z\left(t\right)-s\left(t\right),x^{\prime}-z\left(t\right)-s\left(t\right)\right)\nonumber \\
 & +p_{0}\left(t\right)\left(\frac{dz}{dt}+\frac{ds}{dt}\right)\frac{\partial\psi}{\partial x^{\prime}}\left(x-z\left(t\right)-s\left(t\right),x^{\prime}-z\left(t\right)-s\left(t\right)\right)\nonumber \\
= & -\frac{dp_{0}\left(t\right)}{dt}\psi+p_{0}\left(t\right)\left(\frac{dz}{dt}+\frac{ds}{dt}\right)\left(\frac{\partial\psi}{\partial x}+\frac{\partial\psi}{\partial x^{\prime}}\right)
\end{align}
By combining the terms, 
\begin{align}
\frac{du_{0}\left(t\right)}{dt}E_{4}\left(x-z\left(t\right)\right)-\frac{dz}{dt}u_{0}\left(t\right)\frac{dE_{4}}{dx}\left(x-z\left(t\right)\right) & =F_{u}\left(u,p\right)+\eta\left(x,t\right)\\
-\frac{dp_{0}\left(t\right)}{dt}\psi+p_{0}\left(t\right)\left(\frac{dz}{dt}+\frac{ds}{dt}\right)\left(\frac{\partial\psi}{\partial x}+\frac{\partial\psi}{\partial x^{\prime}}\right) & =F_{p}\left(u,p\right)
\end{align}
By projecting to odd functions respectively, we have
\begin{align}
\frac{dz}{dt}= & -\frac{1}{u_{0}\int dx\left(\frac{dE_{4}}{dx}\right)^{2}}\int dx\frac{dE_{4}}{dx}F_{u}\left(u,p\right)-\frac{1}{u_{0}\int dx\left(\frac{dE_{4}}{dx}\right)^{2}}\int dx\frac{dE_{4}}{dx}\eta\left(x,t\right)\\
\frac{ds}{dt}= & \frac{1}{p_{0}\int dx\int dx^{\prime}\left(\frac{\partial\psi}{\partial x}+\frac{\partial\psi}{\partial x^{\prime}}\right)^{2}}\int dx\int dx^{\prime}\left(\frac{\partial\psi}{\partial x}+\frac{\partial\psi}{\partial x^{\prime}}\right)F_{p}\left(u,p\right)\nonumber \\
 & +\frac{1}{u_{0}\int dx\left(\frac{dE_{4}}{dx}\right)^{2}}\int dx\frac{dE_{4}}{dx}F_{u}\left(u,p\right)+\frac{1}{u_{0}\int dx\left(\frac{dE_{4}}{dx}\right)^{2}}\int dx\frac{dE_{4}}{dx}\eta\left(x,t\right)
\end{align}
By expending $F_{u}$ and $F_{p}$ along $s$, we have
\begin{align}
F_{u}\left[u\left(z\right),p\left(z+s\right)\right]= & F_{u}\left[u\left(z\right),p\left(z\right)\right]+\frac{\partial F_{u}\left[u\left(z\right),p\left(z\right)\right]}{\partial s}s\\
F_{p}\left[u\left(z\right),p\left(z+s\right)\right]= & F_{p}\left[u\left(z\right),p\left(z\right)\right]+\frac{\partial F_{p}\left[u\left(z\right),p\left(z\right)\right]}{\partial s}s.
\end{align}
Then, 
\begin{align}
\frac{dz}{dt}= & M_{zs}s-\tilde{\eta}\left(t\right)\\
\frac{ds}{dt}= & M_{ss}s+\tilde{\eta}\left(t\right),
\end{align}
where
\begin{align}
M_{zs} & =-\frac{1}{u_{0}\int dx\left(\frac{dE_{4}}{dx}\right)^{2}}\int dx\frac{dE_{4}}{dx}\frac{\partial F_{u}\left[u,p\right]}{\partial s}\\
M_{ss} & =\frac{1}{p_{0}\int dx\int dx^{\prime}\left(\frac{\partial\psi}{\partial x}+\frac{\partial\psi}{\partial x^{\prime}}\right)^{2}}\int dx\int dx^{\prime}\left(\frac{\partial\psi}{\partial x}+\frac{\partial\psi}{\partial x^{\prime}}\right)\frac{\partial F_{p}\left[u,p\right]}{\partial s}\nonumber \\
 & \ \ \ \ \ +\frac{1}{u_{0}\int dx\left(\frac{dE_{4}}{dx}\right)^{2}}\int dx\frac{dE_{4}}{dx}\frac{\partial F_{u}\left[u,p\right]}{\partial s}\\
\tilde{\eta}\left(t\right) & =\frac{1}{u_{0}\int dx\left(\frac{dE_{4}}{dx}\right)^{2}}\int dx\frac{dE_{4}}{dx}\eta\left(x,t\right).
\end{align}
Here $\tilde{\eta}\left(t\right)$ is a reduced white noise in the
temporal domain with the following variance.
\begin{align}
\left\langle \tilde{\eta}\left(t\right)\tilde{\eta}\left(t^{\prime}\right)\right\rangle  & =\left(\frac{1}{u_{0}\int dx\left(\frac{dE_{4}}{dx}\right)^{2}}\right)^{2}\int dx_{1}\frac{dE_{4}}{dx_{1}}\int dx_{2}\frac{dE_{4}}{dx_{2}}\left\langle \eta\left(x_{1},t\right)\eta\left(x_{1},t\right)\right\rangle \\
 & =2\frac{\sqrt{8}a}{u_{0}^{2}\sqrt{\pi}}T\delta\left(t-t^{\prime}\right)\\
 & \equiv2\tilde{T}\delta\left(t-t^{\prime}\right)
\end{align}
In a matrix-vector form, we have
\begin{equation}
\frac{d}{dt}\left(\begin{array}{c}
z\\
s
\end{array}\right)=\mathbf{M}\left(\begin{array}{c}
z\\
s
\end{array}\right)+\left(\begin{array}{c}
-1\\
1
\end{array}\right)\tilde{\eta}\left(t\right),
\end{equation}
where
\begin{equation}
\mathbf{M}=\left(\begin{array}{cc}
0 & M_{zs}\\
0 & M_{ss}
\end{array}\right).
\end{equation}
By calculating eigenvalues and eigenvectors of $\mathbf{M}$, we have
\begin{equation}
\frac{d}{dt}U^{-1}\left(\begin{array}{c}
z\\
s
\end{array}\right)=\left(\begin{array}{cc}
E_{0} & 0\\
0 & E_{1}
\end{array}\right)U^{-1}\left(\begin{array}{c}
z\\
s
\end{array}\right)+U^{-1}\left(\begin{array}{c}
-1\\
1
\end{array}\right)\tilde{\eta}\left(t\right),
\end{equation}
where $U=\left(v_{0},v_{1}\right)$ and $E_{n}$s are eigenvalues.
$v_{n}=\left(U_{zn},U_{sn}\right)^{T}$ are corresponding eigenvectors.
By integrating, we have
\begin{align}
\left(\begin{array}{c}
z\\
s
\end{array}\right) & =\int_{0}^{t}dt^{\prime}U\left(\begin{array}{cc}
e^{E_{0}\left(t-t^{\prime}\right)} & 0\\
0 & e^{E_{1}\left(t-t^{\prime}\right)}
\end{array}\right)U^{-1}\left(\begin{array}{c}
-1\\
1
\end{array}\right)\eta\left(x,t^{\prime}\right)\\
 & =\int_{0}^{t}dt^{\prime}\left(\begin{array}{c}
e^{E_{0}\left(t-t^{\prime}\right)}\left(U_{0s}^{-1}-U_{0z}^{-1}\right)U_{z0}+e^{E_{1}\left(t-t^{\prime}\right)}\left(U_{1s}^{-1}-U_{1z}^{-1}\right)U_{z1}\\
e^{E_{0}\left(t-t^{\prime}\right)}\left(U_{0s}^{-1}-U_{0z}^{-1}\right)U_{s0}+e^{E_{1}\left(t-t^{\prime}\right)}\left(U_{1s}^{-1}-U_{1z}^{-1}\right)U_{s1}
\end{array}\right)\tilde{\eta}\left(t\right)
\end{align}
Therefore,
\begin{align}
z= & \int_{0}^{t}dt^{\prime}\left[e^{E_{0}\left(t-t^{\prime}\right)}\left(U_{0s}^{-1}-U_{0z}^{-1}\right)U_{z0}+e^{E_{1}\left(t-t^{\prime}\right)}\left(U_{1s}^{-1}-U_{1z}^{-1}\right)U_{z1}\right]\tilde{\eta}\left(t\right)\\
z^{2}= & \int_{0}^{t}dt^{\prime}\int_{0}^{t}dt^{\prime\prime}\left[e^{E_{0}\left(t-t^{\prime}\right)}\left(U_{0s}^{-1}-U_{0z}^{-1}\right)U_{z0}+e^{E_{1}\left(t-t^{\prime}\right)}\left(U_{1s}^{-1}-U_{1z}^{-1}\right)U_{z1}\right]\nonumber \\
 & \times\left[e^{E_{0}\left(t-t^{\prime\prime}\right)}\left(U_{0s}^{-1}-U_{0z}^{-1}\right)U_{z0}+e^{E_{1}\left(t-t^{\prime\prime}\right)}\left(U_{1s}^{-1}-U_{1z}^{-1}\right)U_{z1}\right]\tilde{\eta}\left(t^{\prime\prime}\right)\tilde{\eta}\left(t^{\prime}\right)\\
\left\langle z^{2}\right\rangle = & 2\tilde{T}\int_{0}^{t}dt^{\prime}\left[e^{E_{0}\left(t-t^{\prime}\right)}\left(U_{0s}^{-1}-U_{0z}^{-1}\right)U_{z0}+e^{E_{1}\left(t-t^{\prime}\right)}\left(U_{1s}^{-1}-U_{1z}^{-1}\right)U_{z1}\right]^{2}
\end{align}
Since $E_{0}=0$, 
\begin{align}
\left\langle z^{2}\right\rangle = & 2\tilde{T}\int_{0}^{t}dt^{\prime}\left[\left(U_{0s}^{-1}-U_{0z}^{-1}\right)U_{z0}+e^{E_{1}\left(t-t^{\prime}\right)}\left(U_{1s}^{-1}-U_{1z}^{-1}\right)U_{z1}\right]^{2}\\
= & 2\tilde{T}\int_{0}^{t}dt^{\prime}\Biggl\{\left[\left(U_{0s}^{-1}-U_{0z}^{-1}\right)U_{z0}\right]^{2}+e^{2E_{1}\left(t-t^{\prime}\right)}\left(U_{1s}^{-1}-U_{1z}^{-1}\right)^{2}U_{z1}^{2}\nonumber \\
 & \ \ \ \ \ \ \ \ +2\left(U_{0s}^{-1}-U_{0z}^{-1}\right)U_{z0}\left(U_{1s}^{-1}-U_{1z}^{-1}\right)U_{z1}e^{E_{1}\left(t-t^{\prime}\right)}\Biggr\}\\
= & 2\tilde{T}\left[\left(U_{0s}^{-1}-U_{0z}^{-1}\right)U_{z0}\right]^{2}t+2\tilde{T}\left(U_{1s}^{-1}-U_{1z}^{-1}\right)^{2}U_{z1}^{2}\frac{1}{2E_{1}}\left(e^{2E_{1}t}-1\right)\nonumber \\
 & \ \ \ +4\tilde{T}\left(U_{0s}^{-1}-U_{0z}^{-1}\right)U_{z0}\left(U_{1s}^{-1}-U_{1z}^{-1}\right)U_{z1}\frac{1}{E_{1}}\left(e^{E_{1}t}-1\right)\\
\approx & 2\tilde{T}\left[\left(U_{0s}^{-1}-U_{0z}^{-1}\right)U_{z0}\right]^{2}t\text{, as \ensuremath{t\rightarrow\infty}}.
\end{align}

To solve it in detail, 
\begin{align}
\det\left(\mathbf{M}-\lambda\mathbf{I}\right) & =\left|\begin{array}{cc}
-\lambda & M_{zs}\\
0 & M_{ss}-\lambda
\end{array}\right|\nonumber \\
 & =-\lambda\left(M_{ss}-\lambda\right)\\
0 & =\det\left(\mathbf{M}-\lambda\mathbf{I}\right)\\
0 & =-\lambda\left(M_{ss}-\lambda\right)
\end{align}
For $\lambda=E_{0}=0$,
\begin{equation}
v_{0}=\left(\begin{array}{c}
1\\
0
\end{array}\right)
\end{equation}
For $\lambda=E_{1}=M_{ss}$,
\begin{align}
\left(\begin{array}{cc}
-M_{ss} & M_{zs}\\
0 & 0
\end{array}\right|\left.\begin{array}{c}
0\\
0
\end{array}\right) & \sim\left(\begin{array}{cc}
-M_{ss} & M_{zs}\end{array}\right|\left.\begin{array}{c}
0\end{array}\right)\\
v_{1} & =\frac{1}{\sqrt{M_{ss}^{2}+M_{zs}^{2}}}\left(\begin{array}{c}
M_{zs}\\
M_{ss}
\end{array}\right)
\end{align}
By grouping terms and solving, we have
\begin{align}
U & =\left(\begin{array}{cc}
1 & \frac{M_{zs}}{\sqrt{M_{ss}^{2}+M_{zs}^{2}}}\\
0 & \frac{M_{ss}}{\sqrt{M_{ss}^{2}+M_{zs}^{2}}}
\end{array}\right)\text{, and}\\
U^{-1} & =\left(\begin{array}{cc}
1 & -\frac{M_{zs}}{M_{ss}}\\
0 & \frac{\sqrt{M_{ss}^{2}+M_{zs}^{2}}}{M_{ss}}
\end{array}\right).
\end{align}
Therefore, the diffusion constant can be solved as 
\begin{align}
D & =2\tilde{T}\left[\left(U_{0s}^{-1}-U_{0z}^{-1}\right)U_{z0}\right]^{2}\\
 & =\frac{4\sqrt{2}a}{u_{0}^{2}\sqrt{\pi}}T\left(1+\frac{M_{zs}}{M_{ss}}\right)^{2}
\end{align}

\bmhead{Acknowledgements}

This work is supported by a Start-up Grant (grant no.: 9610591) for New Faculty and an internal grant (grant no.: 7006055) from the City University of Hong Kong to C.C.A.F. and the general grant (project no.: JCYJ20230807115001004) from the Shenzhen Science and Technology Innovation Committee to C.C.A.F. (the Shenzhen Research Institute, City University of Hong Kong).

\bmhead{Funding}

Open access funding provided by XXX.

\bmhead{Conflict of interest}

Authors declare no conflict of interest of the present work.

\bmhead{Code availability}

The source code of the present study is avaiable at: \url{https://github.com/fccaa}.

\bmhead{Author contribution}
All authors contributed to computer simulations. Z.L and C.C.A.F. contributed manuscript writing. C.C.A.F. contributed mathemtical analysis and project design.

\bibliography{bib_database}


\begin{thebibliography}{30}
\ifx \bisbn   \undefined \def \bisbn  #1{ISBN #1}\fi
\ifx \binits  \undefined \def \binits#1{#1}\fi
\ifx \bauthor  \undefined \def \bauthor#1{#1}\fi
\ifx \batitle  \undefined \def \batitle#1{#1}\fi
\ifx \bjtitle  \undefined \def \bjtitle#1{#1}\fi
\ifx \bvolume  \undefined \def \bvolume#1{\textbf{#1}}\fi
\ifx \byear  \undefined \def \byear#1{#1}\fi
\ifx \bissue  \undefined \def \bissue#1{#1}\fi
\ifx \bfpage  \undefined \def \bfpage#1{#1}\fi
\ifx \blpage  \undefined \def \blpage #1{#1}\fi
\ifx \burl  \undefined \def \burl#1{\textsf{#1}}\fi
\ifx \doiurl  \undefined \def \doiurl#1{\url{https://doi.org/#1}}\fi
\ifx \betal  \undefined \def \betal{\textit{et al.}}\fi
\ifx \binstitute  \undefined \def \binstitute#1{#1}\fi
\ifx \binstitutionaled  \undefined \def \binstitutionaled#1{#1}\fi
\ifx \bctitle  \undefined \def \bctitle#1{#1}\fi
\ifx \beditor  \undefined \def \beditor#1{#1}\fi
\ifx \bpublisher  \undefined \def \bpublisher#1{#1}\fi
\ifx \bbtitle  \undefined \def \bbtitle#1{#1}\fi
\ifx \bedition  \undefined \def \bedition#1{#1}\fi
\ifx \bseriesno  \undefined \def \bseriesno#1{#1}\fi
\ifx \blocation  \undefined \def \blocation#1{#1}\fi
\ifx \bsertitle  \undefined \def \bsertitle#1{#1}\fi
\ifx \bsnm \undefined \def \bsnm#1{#1}\fi
\ifx \bsuffix \undefined \def \bsuffix#1{#1}\fi
\ifx \bparticle \undefined \def \bparticle#1{#1}\fi
\ifx \barticle \undefined \def \barticle#1{#1}\fi
\bibcommenthead
\ifx \bconfdate \undefined \def \bconfdate #1{#1}\fi
\ifx \botherref \undefined \def \botherref #1{#1}\fi
\ifx \url \undefined \def \url#1{\textsf{#1}}\fi
\ifx \bchapter \undefined \def \bchapter#1{#1}\fi
\ifx \bbook \undefined \def \bbook#1{#1}\fi
\ifx \bcomment \undefined \def \bcomment#1{#1}\fi
\ifx \oauthor \undefined \def \oauthor#1{#1}\fi
\ifx \citeauthoryear \undefined \def \citeauthoryear#1{#1}\fi
\ifx \endbibitem  \undefined \def \endbibitem {}\fi
\ifx \bconflocation  \undefined \def \bconflocation#1{#1}\fi
\ifx \arxivurl  \undefined \def \arxivurl#1{\textsf{#1}}\fi
\csname PreBibitemsHook\endcsname

\bibitem[\protect\citeauthoryear{Luo}{2020}]{Luo2020-qp}
\begin{bbook}
\bauthor{\bsnm{Luo}, \binits{L.}}:
\bbtitle{Principles of Neurobiology},
\bedition{2}nd edn.
\bpublisher{CRC Press},
\blocation{London, England}
(\byear{2020})
\end{bbook}
\endbibitem

\bibitem[\protect\citeauthoryear{Tsodyks and Markram}{1997}]{Tsodyks1997}
\begin{barticle}
\bauthor{\bsnm{Tsodyks}, \binits{M.V.}},
\bauthor{\bsnm{Markram}, \binits{H.}}:
\batitle{The neural code between neocortical pyramidal neurons depends on
  neurotransmitter release{\hspace{0.167em}}probability}.
\bjtitle{Proceedings of the National Academy of Sciences}
\bvolume{94}(\bissue{2}),
\bfpage{719}--\blpage{723}
(\byear{1997})
\doiurl{10.1073/pnas.94.2.719}
\end{barticle}
\endbibitem

\bibitem[\protect\citeauthoryear{Grillo et~al.}{2018}]{Grillo2018}
\begin{barticle}
\bauthor{\bsnm{Grillo}, \binits{F.W.}},
\bauthor{\bsnm{Neves}, \binits{G.}},
\bauthor{\bsnm{Walker}, \binits{A.}},
\bauthor{\bsnm{Vizcay-Barrena}, \binits{G.}},
\bauthor{\bsnm{Fleck}, \binits{R.A.}},
\bauthor{\bsnm{Branco}, \binits{T.}},
\bauthor{\bsnm{Burrone}, \binits{J.}}:
\batitle{A distance-dependent distribution of presynaptic boutons tunes
  frequency-dependent dendritic integration}.
\bjtitle{Neuron}
\bvolume{99}(\bissue{2}),
\bfpage{275}--\blpage{2823}
(\byear{2018})
\doiurl{10.1016/j.neuron.2018.06.015}
\end{barticle}
\endbibitem

\bibitem[\protect\citeauthoryear{Jensen et~al.}{2021}]{Jensen2021}
\begin{botherref}
\oauthor{\bsnm{Jensen}, \binits{T.P.}},
\oauthor{\bsnm{Kopach}, \binits{O.}},
\oauthor{\bsnm{Reynolds}, \binits{J.P.}},
\oauthor{\bsnm{Savtchenko}, \binits{L.P.}},
\oauthor{\bsnm{Rusakov}, \binits{D.A.}}:
Release probability increases towards distal dendrites boosting high-frequency
  signal transfer in the rodent hippocampus.
{eLife}
\textbf{10}
(2021)
\doiurl{10.7554/elife.62588}
\end{botherref}
\endbibitem

\bibitem[\protect\citeauthoryear{Chipman et~al.}{2021}]{Chipman2021}
\begin{botherref}
\oauthor{\bsnm{Chipman}, \binits{P.H.}},
\oauthor{\bsnm{Fung}, \binits{C.C.A.}},
\oauthor{\bsnm{Fernandez}, \binits{A.P.}},
\oauthor{\bsnm{Sawant}, \binits{A.}},
\oauthor{\bsnm{Tedoldi}, \binits{A.}},
\oauthor{\bsnm{Kawai}, \binits{A.}},
\oauthor{\bsnm{Ghimire~Gautam}, \binits{S.}},
\oauthor{\bsnm{Kurosawa}, \binits{M.}},
\oauthor{\bsnm{Abe}, \binits{M.}},
\oauthor{\bsnm{Sakimura}, \binits{K.}},
\oauthor{\bsnm{Fukai}, \binits{T.}},
\oauthor{\bsnm{Goda}, \binits{Y.}}:
Astrocyte {GluN}2c {NMDA} receptors control basal synaptic strengths of
  hippocampal {CA}1 pyramidal neurons in the stratum radiatum.
{eLife}
\textbf{10}
(2021)
\doiurl{10.7554/elife.70818}
\end{botherref}
\endbibitem

\bibitem[\protect\citeauthoryear{Zorumski et~al.}{2015}]{Zorumski2015}
\begin{botherref}
\oauthor{\bsnm{Zorumski}, \binits{C.F.}},
\oauthor{\bsnm{Nagele}, \binits{P.}},
\oauthor{\bsnm{Mennerick}, \binits{S.}},
\oauthor{\bsnm{Conway}, \binits{C.R.}}:
Treatment-resistant major depression: Rationale for nmda receptors as targets
  and nitrous oxide as therapy.
Frontiers in Psychiatry
\textbf{6}
(2015)
\doiurl{10.3389/fpsyt.2015.00172}
\end{botherref}
\endbibitem

\bibitem[\protect\citeauthoryear{Williams and Schatzberg}{2016}]{Williams2016}
\begin{barticle}
\bauthor{\bsnm{Williams}, \binits{N.R.}},
\bauthor{\bsnm{Schatzberg}, \binits{A.F.}}:
\batitle{{NMDA} antagonist treatment of depression}.
\bjtitle{Current Opinion in Neurobiology}
\bvolume{36},
\bfpage{112}--\blpage{117}
(\byear{2016})
\doiurl{10.1016/j.conb.2015.11.001}
\end{barticle}
\endbibitem

\bibitem[\protect\citeauthoryear{Krystal et~al.}{2019}]{Krystal2019}
\begin{barticle}
\bauthor{\bsnm{Krystal}, \binits{J.H.}},
\bauthor{\bsnm{Abdallah}, \binits{C.G.}},
\bauthor{\bsnm{Sanacora}, \binits{G.}},
\bauthor{\bsnm{Charney}, \binits{D.S.}},
\bauthor{\bsnm{Duman}, \binits{R.S.}}:
\batitle{Ketamine: A paradigm shift for depression research and treatment}.
\bjtitle{Neuron}
\bvolume{101}(\bissue{5}),
\bfpage{774}--\blpage{778}
(\byear{2019})
\doiurl{10.1016/j.neuron.2019.02.005}
\end{barticle}
\endbibitem

\bibitem[\protect\citeauthoryear{Amari}{1977}]{Amari1977}
\begin{barticle}
\bauthor{\bsnm{Amari}, \binits{S.-i.}}:
\batitle{Dynamics of pattern formation in lateral-inhibition type neural
  fields}.
\bjtitle{Biological Cybernetics}
\bvolume{27}(\bissue{2}),
\bfpage{77}--\blpage{87}
(\byear{1977})
\doiurl{10.1007/bf00337259}
\end{barticle}
\endbibitem

\bibitem[\protect\citeauthoryear{Georgopoulos et~al.}{1993}]{Georgopoulos1993}
\begin{barticle}
\bauthor{\bsnm{Georgopoulos}, \binits{A.P.}},
\bauthor{\bsnm{Taira}, \binits{M.}},
\bauthor{\bsnm{Lukashin}, \binits{A.}}:
\batitle{Cognitive neurophysiology of the motor cortex}.
\bjtitle{Science}
\bvolume{260}(\bissue{5104}),
\bfpage{47}--\blpage{52}
(\byear{1993})
\doiurl{10.1126/science.8465199}
\end{barticle}
\endbibitem

\bibitem[\protect\citeauthoryear{Ben-Yishai et~al.}{1995}]{BenYishai1995}
\begin{barticle}
\bauthor{\bsnm{Ben-Yishai}, \binits{R.}},
\bauthor{\bsnm{Lev~Bar-Or}, \binits{R.}},
\bauthor{\bsnm{Sompolinsky}, \binits{H.}}:
\batitle{Theory of orientation tuning in visual cortex.}
\bjtitle{Proceedings of the National Academy of Sciences}
\bvolume{92}(\bissue{9}),
\bfpage{3844}--\blpage{3848}
(\byear{1995})
\doiurl{10.1073/pnas.92.9.3844}
\end{barticle}
\endbibitem

\bibitem[\protect\citeauthoryear{Fung et~al.}{2010}]{Fung2010}
\begin{barticle}
\bauthor{\bsnm{Fung}, \binits{C.C.A.}},
\bauthor{\bsnm{Wong}, \binits{K.Y.M.}},
\bauthor{\bsnm{Wu}, \binits{S.}}:
\batitle{A moving bump in a continuous manifold: A comprehensive study of the
  tracking dynamics of continuous attractor neural networks}.
\bjtitle{Neural Computation}
\bvolume{22}(\bissue{3}),
\bfpage{752}--\blpage{792}
(\byear{2010})
\doiurl{10.1162/neco.2009.07-08-824}
\end{barticle}
\endbibitem

\bibitem[\protect\citeauthoryear{Zhang}{1996}]{Zhang1996}
\begin{barticle}
\bauthor{\bsnm{Zhang}, \binits{K.}}:
\batitle{Representation of spatial orientation by the intrinsic dynamics of the
  head-direction cell ensemble: a theory}.
\bjtitle{The Journal of Neuroscience}
\bvolume{16}(\bissue{6}),
\bfpage{2112}--\blpage{2126}
(\byear{1996})
\doiurl{10.1523/jneurosci.16-06-02112.1996}
\end{barticle}
\endbibitem

\bibitem[\protect\citeauthoryear{Battaglia and Treves}{1998}]{Battaglia1998}
\begin{barticle}
\bauthor{\bsnm{Battaglia}, \binits{F.P.}},
\bauthor{\bsnm{Treves}, \binits{A.}}:
\batitle{Attractor neural networks storing multiple space representations: A
  model for hippocampal place fields}.
\bjtitle{Physical Review E}
\bvolume{58}(\bissue{6}),
\bfpage{7738}--\blpage{7753}
(\byear{1998})
\doiurl{10.1103/physreve.58.7738}
\end{barticle}
\endbibitem

\bibitem[\protect\citeauthoryear{Markram et~al.}{1998}]{Markram1998}
\begin{barticle}
\bauthor{\bsnm{Markram}, \binits{H.}},
\bauthor{\bsnm{Wang}, \binits{Y.}},
\bauthor{\bsnm{Tsodyks}, \binits{M.}}:
\batitle{Differential signaling via the same axon of neocortical pyramidal
  neurons}.
\bjtitle{Proceedings of the National Academy of Sciences}
\bvolume{95}(\bissue{9}),
\bfpage{5323}--\blpage{5328}
(\byear{1998})
\doiurl{10.1073/pnas.95.9.5323}
\end{barticle}
\endbibitem

\bibitem[\protect\citeauthoryear{Wang et~al.}{2015}]{Wang2015}
\begin{botherref}
\oauthor{\bsnm{Wang}, \binits{H.}},
\oauthor{\bsnm{Lam}, \binits{K.}},
\oauthor{\bsnm{Fung}, \binits{C.C.A.}},
\oauthor{\bsnm{Wong}, \binits{K.Y.M.}},
\oauthor{\bsnm{Wu}, \binits{S.}}:
Rich spectrum of neural field dynamics in the presence of short-term synaptic
  depression.
Physical Review E
\textbf{92}(3)
(2015)
\doiurl{10.1103/physreve.92.032908}
\end{botherref}
\endbibitem

\bibitem[\protect\citeauthoryear{Stein et~al.}{2021}]{Stein2021}
\begin{barticle}
\bauthor{\bsnm{Stein}, \binits{H.}},
\bauthor{\bsnm{Barbosa}, \binits{J.}},
\bauthor{\bsnm{Compte}, \binits{A.}}:
\batitle{Towards biologically constrained attractor models of schizophrenia}.
\bjtitle{Current Opinion in Neurobiology}
\bvolume{70},
\bfpage{171}--\blpage{181}
(\byear{2021})
\doiurl{10.1016/j.conb.2021.10.013}
\end{barticle}
\endbibitem

\bibitem[\protect\citeauthoryear{Fung et~al.}{2012}]{Fung2012a}
\begin{barticle}
\bauthor{\bsnm{Fung}, \binits{C.C.A.}},
\bauthor{\bsnm{Wong}, \binits{K.Y.M.}},
\bauthor{\bsnm{Wang}, \binits{H.}},
\bauthor{\bsnm{Wu}, \binits{S.}}:
\batitle{Dynamical synapses enhance neural information processing:
  Gracefulness, accuracy, and mobility}.
\bjtitle{Neural Computation}
\bvolume{24}(\bissue{5}),
\bfpage{1147}--\blpage{1185}
(\byear{2012})
\doiurl{10.1162/neco_a_00269}
\end{barticle}
\endbibitem

\bibitem[\protect\citeauthoryear{Wu and Amari}{2005}]{Wu2005}
\begin{barticle}
\bauthor{\bsnm{Wu}, \binits{S.}},
\bauthor{\bsnm{Amari}, \binits{S.-i.}}:
\batitle{Computing with continuous attractors: stability and online aspects}.
\bjtitle{Neural computation}
\bvolume{17}(\bissue{10}),
\bfpage{2215}--\blpage{2239}
(\byear{2005})
\doiurl{10.1162/0899766054615626}
\end{barticle}
\endbibitem

\bibitem[\protect\citeauthoryear{Deneve et~al.}{1999}]{Deneve1999}
\begin{barticle}
\bauthor{\bsnm{Deneve}, \binits{S.}},
\bauthor{\bsnm{Latham}, \binits{P.E.}},
\bauthor{\bsnm{Pouget}, \binits{A.}}:
\batitle{Reading population codes: a neural implementation of ideal observers}.
\bjtitle{Nature Neuroscience}
\bvolume{2}(\bissue{8}),
\bfpage{740}--\blpage{745}
(\byear{1999})
\doiurl{10.1038/11205}
\end{barticle}
\endbibitem

\bibitem[\protect\citeauthoryear{Tsodyks et~al.}{1998}]{Tsodyks1998}
\begin{barticle}
\bauthor{\bsnm{Tsodyks}, \binits{M.}},
\bauthor{\bsnm{Pawelzik}, \binits{K.}},
\bauthor{\bsnm{Markram}, \binits{H.}}:
\batitle{Neural networks with dynamic synapses}.
\bjtitle{Neural Computation}
\bvolume{10}(\bissue{4}),
\bfpage{821}--\blpage{835}
(\byear{1998})
\doiurl{10.1162/089976698300017502}
\end{barticle}
\endbibitem

\bibitem[\protect\citeauthoryear{Cochilla et~al.}{1999}]{Cochilla1999}
\begin{barticle}
\bauthor{\bsnm{Cochilla}, \binits{A.J.}},
\bauthor{\bsnm{Angleson}, \binits{J.K.}},
\bauthor{\bsnm{Betz}, \binits{W.J.}}:
\batitle{Monitoring secretory membrane with fm1-43 fluorescence}.
\bjtitle{Annual Review of Neuroscience}
\bvolume{22}(\bissue{1}),
\bfpage{1}--\blpage{10}
(\byear{1999})
\doiurl{10.1146/annurev.neuro.22.1.1}
\end{barticle}
\endbibitem

\bibitem[\protect\citeauthoryear{Wu et~al.}{2008}]{Wu2008}
\begin{barticle}
\bauthor{\bsnm{Wu}, \binits{S.}},
\bauthor{\bsnm{Hamaguchi}, \binits{K.}},
\bauthor{\bsnm{Amari}, \binits{S.-i.}}:
\batitle{{Dynamics and Computation of Continuous Attractors}}.
\bjtitle{Neural Computation}
\bvolume{20}(\bissue{4}),
\bfpage{994}--\blpage{1025}
(\byear{2008})
\doiurl{10.1162/neco.2008.10-06-378}
{\href{https://arxiv.org/abs/https://direct.mit.edu/neco/article-pdf/20/4/994/817269/neco.2008.10-06-378.pdf}{{https://direct.mit.edu/neco/article-pdf/20/4/994/817269/neco.2008.10-06-378.pdf}}}
\end{barticle}
\endbibitem

\bibitem[\protect\citeauthoryear{Paoletti et~al.}{2013}]{Paoletti2013}
\begin{barticle}
\bauthor{\bsnm{Paoletti}, \binits{P.}},
\bauthor{\bsnm{Bellone}, \binits{C.}},
\bauthor{\bsnm{Zhou}, \binits{Q.}}:
\batitle{{NMDA} receptor subunit diversity: impact on receptor properties,
  synaptic plasticity and disease}.
\bjtitle{Nat. Rev. Neurosci.}
\bvolume{14}(\bissue{6}),
\bfpage{383}--\blpage{400}
(\byear{2013})
\end{barticle}
\endbibitem

\bibitem[\protect\citeauthoryear{Parsons et~al.}{1998}]{Parsons1998}
\begin{barticle}
\bauthor{\bsnm{Parsons}, \binits{C.G.}},
\bauthor{\bsnm{Danysz}, \binits{W.}},
\bauthor{\bsnm{Quack}, \binits{G.}}:
\batitle{Glutamate in cns disorders as a target for drug development: an
  update}.
\bjtitle{Drug news \& perspectives}
\bvolume{11}(\bissue{9}),
\bfpage{523}--\blpage{569}
(\byear{1998})
\doiurl{10.1358/dnp.1998.11.9.863689}
\end{barticle}
\endbibitem

\bibitem[\protect\citeauthoryear{Zarate et~al.}{2006}]{Zarate2006}
\begin{barticle}
\bauthor{\bsnm{Zarate}, \binits{C.A.} \bsuffix{Jr}},
\bauthor{\bsnm{Singh}, \binits{J.B.}},
\bauthor{\bsnm{Carlson}, \binits{P.J.}},
\bauthor{\bsnm{Brutsche}, \binits{N.E.}},
\bauthor{\bsnm{Ameli}, \binits{R.}},
\bauthor{\bsnm{Luckenbaugh}, \binits{D.A.}},
\bauthor{\bsnm{Charney}, \binits{D.S.}},
\bauthor{\bsnm{Manji}, \binits{H.K.}}:
\batitle{A randomized trial of an {N-methyl-D-aspartate} antagonist in
  treatment-resistant major depression}.
\bjtitle{Arch. Gen. Psychiatry}
\bvolume{63}(\bissue{8}),
\bfpage{856}--\blpage{864}
(\byear{2006})
\end{barticle}
\endbibitem

\bibitem[\protect\citeauthoryear{Hardingham and Bading}{2010}]{Hardingham2010}
\begin{barticle}
\bauthor{\bsnm{Hardingham}, \binits{G.E.}},
\bauthor{\bsnm{Bading}, \binits{H.}}:
\batitle{Synaptic versus extrasynaptic {NMDA} receptor signalling: implications
  for neurodegenerative disorders}.
\bjtitle{Nat. Rev. Neurosci.}
\bvolume{11}(\bissue{10}),
\bfpage{682}--\blpage{696}
(\byear{2010})
\end{barticle}
\endbibitem

\bibitem[\protect\citeauthoryear{Muir and Lees}{1995}]{Muir1995}
\begin{barticle}
\bauthor{\bsnm{Muir}, \binits{K.W.}},
\bauthor{\bsnm{Lees}, \binits{K.R.}}:
\batitle{Clinical experience with excitatory amino acid antagonist drugs}.
\bjtitle{Stroke}
\bvolume{26}(\bissue{3}),
\bfpage{503}--\blpage{513}
(\byear{1995})
\end{barticle}
\endbibitem

\bibitem[\protect\citeauthoryear{Fung et~al.}{2012}]{Fung2012b}
\begin{bchapter}
\bauthor{\bsnm{Fung}, \binits{C.C.A.}},
\bauthor{\bsnm{Wong}, \binits{K.Y.M.}},
\bauthor{\bsnm{Wu}, \binits{S.}}:
\bctitle{Delay compensation with dynamical synapses}.
In: \beditor{\bsnm{Pereira}, \binits{F.}},
\beditor{\bsnm{Burges}, \binits{C.J.}},
\beditor{\bsnm{Bottou}, \binits{L.}},
\beditor{\bsnm{Weinberger}, \binits{K.Q.}} (eds.)
\bbtitle{Advances in Neural Information Processing Systems},
vol. \bseriesno{25}.
\bpublisher{Curran Associates, Inc.}, \blocation{???}
(\byear{2012}).
\burl{https://proceedings.neurips.cc/paper/2012/file/85422afb467e9456013a2a51d4dff702-Paper.pdf}
\end{bchapter}
\endbibitem

\bibitem[\protect\citeauthoryear{Diazgranados et~al.}{2010}]{Diazgranados2010}
\begin{barticle}
\bauthor{\bsnm{Diazgranados}, \binits{N.}},
\bauthor{\bsnm{Ibrahim}, \binits{L.}},
\bauthor{\bsnm{Brutsche}, \binits{N.E.}},
\bauthor{\bsnm{Newberg}, \binits{A.}},
\bauthor{\bsnm{Kronstein}, \binits{P.}},
\bauthor{\bsnm{Khalife}, \binits{S.}},
\bauthor{\bsnm{Kammerer}, \binits{W.A.}},
\bauthor{\bsnm{Quezado}, \binits{Z.}},
\bauthor{\bsnm{Luckenbaugh}, \binits{D.A.}},
\bauthor{\bsnm{Salvadore}, \binits{G.}},
\bauthor{\bsnm{Machado-Vieira}, \binits{R.}},
\bauthor{\bsnm{Manji}, \binits{H.K.}},
\bauthor{\bsnm{Zarate}, \binits{J.} \bsuffix{Carlos~A.}}:
\batitle{{A Randomized Add-on Trial of an N-methyl-D-aspartate Antagonist in
  Treatment-Resistant Bipolar Depression}}.
\bjtitle{Archives of General Psychiatry}
\bvolume{67}(\bissue{8}),
\bfpage{793}--\blpage{802}
(\byear{2010})
\doiurl{10.1001/archgenpsychiatry.2010.90}
{\href{https://arxiv.org/abs/https://jamanetwork.com/journals/jamapsychiatry/articlepdf/210856/yoa05010\_793\_802.pdf}{{https://jamanetwork.com/journals/jamapsychiatry/articlepdf/210856/yoa05010\_793\_802.pdf}}}
\end{barticle}
\endbibitem

\end{thebibliography}

\end{document}